\documentclass[aps,prl,article,twocolumn,floatfix,superscriptaddress,longbibliography]{revtex4-2}
\usepackage{graphicx}
\usepackage{epstopdf}
\usepackage{dcolumn}
\usepackage{bm}
\usepackage{physics}
\usepackage{xcolor}
\usepackage{amssymb}
\usepackage{amsmath}
\usepackage{bbold}
\usepackage{soul}
\usepackage[colorlinks=true, linkcolor=blue, urlcolor=blue, citecolor=blue]{hyperref}
\usepackage{soul}
\usepackage[normalem]{ulem}

\newcommand*{\la}{\langle}
\newcommand*{\ra}{\rangle}

\newcommand{\affFUW}{Faculty of Physics, University of Warsaw, Pasteura 5, 02-093 Warsaw, Poland}
\newcommand{\affWroclaw}{Institute of Theoretical Physics, Wrocław University of Science and Technology, 50-370 Wrocław, Poland}
\newcommand{\affICFO}{ICFO-Institut de Ciencies Fotoniques, The Barcelona Institute of Science and Technology, 08860 Castelldefels (Barcelona), Spain}
\newcommand{\affDIPC}{DIPC - Donostia International Physics Center, Paseo Manuel de Lardiz{\'a}bal 4, 20018 San Sebasti{\'a}n, Spain}
\newcommand{\affIker}{IKERBASQUE, Basque Foundation for Science, Plaza Euskadi 5, 48009 Bilbao, Spain}
\newcommand{\affCent}{Centre for Quantum Optical Technologies, Centre of New Technologies, University of Warsaw, Banacha 2c, 02-097 Warsaw, Poland}

\begin{document}
\preprint{APS/123-QED}

\title{Controlled pairing symmetries in a Fermi-Hubbard ladder with band flattening}

\author{Jo\~{a}o P. Mendon\c{c}a} 
\email{jpedromend@gmail.com}
\affiliation{\affCent}
\affiliation{\affFUW}
\author{S. Biswas}
\affiliation{\affDIPC}
\author{M. Dziurawiec}
\affiliation{\affWroclaw}
\author{U. Bhattacharya}
\affiliation{Institute for Theoretical Physics, ETH Z\"urich, Z\"urich 8093, Switzerland}
\author{K.~Jachymski}
\affiliation{\affFUW}
\author{M. Aidelsburger}
\affiliation{Max-Planck-Institut f\"ur Quantenoptik, 85748 Garching, Germany}
\affiliation{Munich Center for Quantum Science and Technology (MCQST), 80799 M\"unchen, Germany}
\affiliation{Fakult\"at f\"ur Physik, Ludwig-Maximilians-Universit\"at, 80799 M\"unchen, Germany}
\author{M. Lewenstein}
\affiliation{\affICFO}
\author{M. M. Maśka}
\affiliation{\affWroclaw}
\author{T. Grass}
\email{tobias.grass@dipc.org}
\affiliation{\affDIPC}
\affiliation{\affIker}

\date{\today}

\begin{abstract}
Band flattening has been identified as key ingredient to correlation phenomena in Moiré materials and beyond. 
Here, we examine strongly repulsive fermions on a ladder -- a minimal platform for unconventional $d$-wave pairing -- and show that flattening of the lower band through an additional diagonal hopping term produces non-Fermi liquid behavior, evidenced by the violation of Luttinger's theorem, as well as axial $d$-wave pairing correlations. Alternatively, plaquette ring exchange can also generate pairing, albeit with a distinct diagonal $d$-wave pairing symmetry. Hence, our finding showcases a competition of different unconventional pairing channels, and demonstrates via a simple model how band geometry can induce fermionic pairing. This offers broadly relevant insights for correlated flat-band systems, ranging from ultracold atoms to strongly interacting electrons in solids.
\end{abstract}

\maketitle
\begin{figure}[t]
    \centering
    \includegraphics[width=0.9\columnwidth]{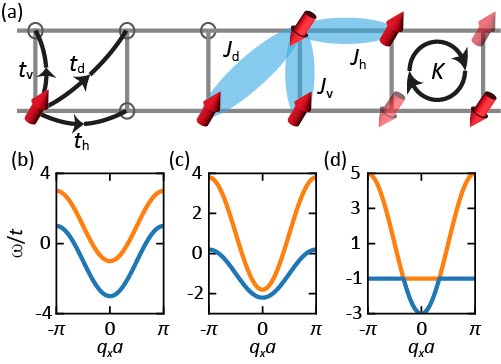}
    \caption{(a) Schematic representation of the extended $t-J-K$ model [cf. Eq.~\eqref{eq:model}]. Electrons are allowed to hop between all neighboring sites, including horizontal, vertical, and diagonals. The exchange interaction is tied to the hoppings via $J=4t^2/U$. The ring-exchange interaction acting on a single plaquette is shown. (b-d): Single-particle bands for (b) $t_d=0$, (c) $t_d=-0.4$, and (d) $t_d=-1$ (in units of $t=t_h=t_v$). 
    }
    \label{fig:sketch}
\end{figure}

\textit{Introduction.} Since the discovery of twisted bilayer graphene at the magic angle \cite{Cao2018}, flat energy bands have become a central paradigm in the search for strongly correlated phases, both in electronic systems \cite{Wang2020,Zhang2020,Uri2023,Li2024,Hao2024} and in synthetic quantum matter \cite{Gonzalez-Tudela2019,Salamon2020,Salamon2022,Meng2023}; see Ref.~\onlinecite{Torma2022} for a review. While twisting offers a powerful route to band flattening, geometric control in single-layer systems provides an alternative and conceptually simpler mechanism. Prominent examples include the Lieb and Kagome lattices, where flat bands arise naturally and have stimulated extensive studies of flat-band Bose--Einstein condensation \cite{You2012,Julku2021,Jalali-mola2023}.
From a theoretical standpoint, one of the simplest settings with tunable bandwidth is a two-leg ladder in square geometry augmented by a diagonal next-nearest-neighbor hopping $t_d$, cf.~Fig.~\ref{fig:sketch}(a). The resulting band structure consists of two bands with bandwidths $2|t_h\pm t_d|$, where $t_h$ denotes hopping along the legs. By tuning $t_d$, one band can be continuously flattened and even rendered exactly dispersionless, as shown in Figs.~\ref{fig:sketch}(b)--(d). This construction offers a minimal and highly controllable route to flat-band physics.
Crucially, the two-leg ladder is also the minimal geometry that extends one-dimensional systems into a second dimension, thereby permitting spatially nontrivial pairing correlations. In particular, it naturally supports $d$-wave symmetry. Jiang \textit{et al.}~\cite{jiang2013non} demonstrated that robust $d$-wave correlations can be stabilized in this geometry by introducing a ring-exchange interaction $K$, which coherently transfers an electronic singlet between the two diagonals of a plaquette. Related bosonic studies revealed analogous $d$-wave correlations and identified the ladder as a promising route toward Bose-metal behavior \cite{Motrunich_2007,Sheng2008,Mishmash_2011,Biswas2025_2}, an elusive non-condensed phase with a recently proposed realization in trapped-ion platforms \cite{Biswas2025}.
In the fermionic case—where the state may be viewed as a product of a bosonic $d$-wave phase of chargons and a Slater determinant of spinons \cite{jiang2013non}—the emergence of $d$-wave pairing is accompanied by a fundamental reconstruction of the electronic structure. Most notably, Luttinger’s theorem is violated \cite{Luttinger1960}, signaling the breakdown of the conventional Fermi sea and the onset of a non–Fermi-liquid (NFL) phase. However, access to this regime requires a ring-exchange strength comparable to the bare hopping, posing a practical limitation.
This observation motivates the central idea of this work: combining ring-exchange interactions with band flattening to amplify correlation effects.

In this Letter, we develop this idea by studying a Fermi--Hubbard ladder with diagonal hopping $t_d$ and ring-exchange interactions $K$. For $0>t_d>-0.5t_h$, the width of the lower band is reduced while a finite band gap is preserved, see Fig.~\ref{fig:sketch}(c). Focusing on this regime, we show that diagonal hopping provides a natural and experimentally relevant knob to enhance pairing tendencies and stabilize non--Fermi-liquid (NFL) behavior at significantly reduced interaction strengths.
Moreover, we find that the NFL regime splits into two distinct phases, most clearly distinguished by their $d$-wave pairing symmetries. While ring-exchange interactions favor diagonal $d_{xy}$-wave pairing, diagonal hopping promotes axial $d_{x^2-y^2}$-wave correlations. Our results thus reveal a direct competition between interaction-driven and geometry-induced pairing mechanisms and establish, within a minimal model, how band flattening via diagonal hopping enhances fermionic pairing in ladder systems.
These findings are directly relevant for experiments with ultracold atoms \cite{Hirthe2023}, quantum dot arrays \cite{Hsiao2024}, and quasi-one-dimensional, multi-orbital strongly correlated materials \cite{Herbrych2018,Herbrych2019}. 
We further note that artificial gauge fields have achieved complex or negative tunneling amplitudes in optical lattice experiments, including systems with synthetic ladder geometry  \cite{Atala2014,Stuhl2015,Mancini2015, Tai2017, song_observation_2018, kang_creutz_2020, Zhou2023,Impertro2025}. More broadly, our work provides a transparent example of how band flattening can qualitatively reshape correlated many-body phases, even in simple lattice geometries.

\textit{Model.} Following earlier work on NFL $d$-wave phase~\cite{jiang2013non}, we assume a strongly repulsive Fermi-Hubbard system that is faithfully described by the $t$--$J$ model, disregarding states with doubly occupied sites, thereby limiting the interactions to a second-order exchange process $J_{ij}=4t_{ij}^2/U$, where $i,j$ are site indices, and the interaction is strong, $U\gg t$. As indicated in Fig.~\ref{fig:sketch}(a), we assume non-zero hopping terms ($t_h$, $t_v$, $t_d$) as well as non-zero exchange terms ($J_h$, $J_v$, $J_d$) for nearest neighbors along the leg and rung, as well as next-nearest neighbors along the diagonals of each plaquette. 
Throughout this paper, we set $U=2$ and $t_h=t_v=t=1$, while $t_d$ is kept tunable. 
While we focus on the two-leg ladder structure, the model has also received attention in the 2D context~\cite{Gong2021,Jiang2021}, where pairing tendencies can condense into true long-range superconducting order. In addition to hopping and exchange terms, we also include a tunable ring exchange term $K$. The total $t$--$J$--$K$ Hamiltonian reads:
\begin{align}\label{eq:model}
    H &= -\sum_{\la i,j\ra,\sigma}t_{ij}(\hat{c}^\dagger_{i\sigma} \hat{c}_{j\sigma} + \textrm{H.c.}) 
    + \sum_{\la i,j \ra} J_{ij} \hat{\bm{S}}_i\cdot\hat{\bm{S}}_j
    \nonumber\\
    &+ 2K \sum_{\square} ( \hat{\Delta}_{13}^\dagger \hat{\Delta}_{24} + \textrm{H.c.}),
\end{align}
where $\hat{\Delta}^\dagger_{ij}=(\hat{c}_{i \uparrow}^\dagger \hat{c}_{j \downarrow}^\dagger - \hat{c}_{i \downarrow}^\dagger \hat{c}_{j \uparrow}^\dagger)/\sqrt{2}$. The spin operators are defined as $\hat{\bm{S}}_i=\frac{1}{2}\sum_{\sigma,\sigma'} \hat{c}_{i\sigma}^\dagger \bm{\sigma}_{\sigma\sigma'} \hat{c}_{i\sigma'}$. The sums are carried out over nearest-neighbor sites $\la i,j \ra$, and along  elementary plaquettes $\square$.
For simplicity, and as done in earlier work~\cite{jiang2013non}, we have neglected the $-J \hat{n}_i \hat{n}_j/4$ term of the standard $t$--$J$ model.

\begin{figure}[t]
    \centering
    \includegraphics[width=0.9\columnwidth]{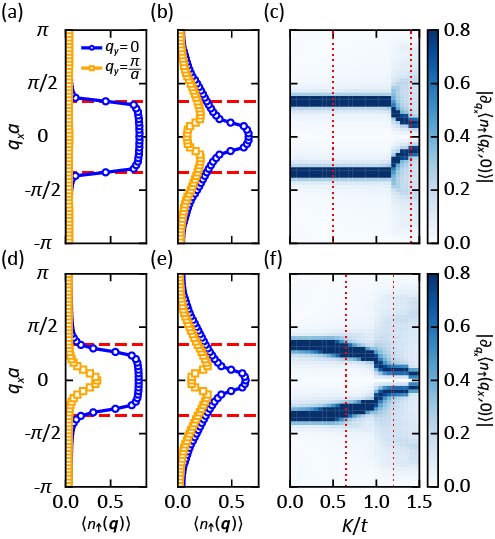}
    \caption{(a,b,d,e) Momentum distribution $\la \hat{n}_{\uparrow}(\bm{q}) \ra$ for $L_x=30$, and fixed (a-c) $t_d=-0.1$ and (d-f) $t_d=-0.3$. (a) Shows FL behavior at $K=0.5$, whereas panel (b) NFL behavior at $K=1.4$. Both (d) at $K=0.65$ and (e) at $K=1.2$ showcase distinct NFL behavior. 
    In panels (c,f) the absolute value of the derivative $\partial \la \hat{n}_{\uparrow}(q_x,0) \ra/\partial q_x$ is plotted as a function of $q_x$ and $K$ for $q_y=0$. Dotted lines in (c,f) mark the $K$ values which are shown in panels (a,b,d,e). The Fermi wave vector $k_F$ is represented by horizontal red dashed lines. 
    }
    \label{fig:moment}
\end{figure}

\textit{Non-Fermi liquid phases.}
Using DMRG calculations \footnote{The numerical simulations in this work were performed using the ITensors library in Julia \cite{ITensor,ITensor-r0.3}. The energy convergence tolerance was set to $10^{-10}$, and results were recorded only after reaching this precision. The truncation error was controlled by a cutoff of $10^{-10}$, yielding a maximum truncation error of approximately $10^{-10}$ in the final DMRG sweeps.},
we first analyze the momentum distribution
$\langle n_{\uparrow}(\bm{q}) \rangle = \langle \hat{c}_{\bm{q},\uparrow}^\dagger \hat{c}_{\bm{q},\uparrow} \rangle$.
In a Fermi liquid (FL), the momentum distribution obeys the Luttinger liquid (LL) theorem and exhibits a singular surface associated with the Fermi sea, with a non-analyticity at the Fermi wavevector $k_F=\pi\rho$. The particle density is $\rho=N_e/N$, where $N_e=N_\uparrow+N_\downarrow$ and $N=2L_x$. Throughout this work, we fix $\rho=1/3$.
The FL momentum distribution is shown in Fig.~\ref{fig:moment}(a) for $t_d=-0.1$ and $K=0.5$, where only modes with $q_y=0$ and $|q_x|\lesssim k_F$ are occupied. Increasing $K$ drives a sharp qualitative change in $\langle n_{\uparrow}(\bm{q}) \rangle$ as the system enters a NFL regime. As illustrated in Fig.~\ref{fig:moment}(b) for $t_d=-0.1$ and $K=1.4$, the momentum distribution no longer displays a well-defined Fermi surface, signaling a strong breakdown of the LL theorem.
The nature of the FL--NFL transition is made explicit by the derivative $|\partial_{q_x}\langle n_{\uparrow}(\bm{q}) \rangle|$, shown in Fig.~\ref{fig:moment}(c) as a function of $K$. In the FL phase, the dominant peaks are pinned to $|q_x|=k_F$. By contrast, in the $d_{xy}$-NFL regime these peaks shift abruptly to incommensurate momenta, providing a clear and unambiguous signature of a strong violation of the LL theorem.
At stronger diagonal hopping, $t_d=-0.3$, the violation of the LL theorem takes a qualitatively different form. As shown in Fig.~\ref{fig:moment}(f), deviations from FL behavior emerge already for $K\gtrsim0.5$, but now through a continuous evolution of the peak positions rather than an abrupt jump. The corresponding momentum distribution at intermediate coupling, Fig.~\ref{fig:moment}(d) for $K=0.65$, exhibits a weak population of the $q_y=\pi$ sector, $0<\langle n_{\uparrow}(q_x,\pi)\rangle<1$ at small $q_x$, together with a reduced $q_y=0$ plateau terminating at $q_x<k_F$. This smooth redistribution of spectral weight indicates a weak violation of the LL theorem.
Upon further increasing $K$, the system crosses over into a regime with pronounced NFL character. As shown in Fig.~\ref{fig:moment}(e) for $K=1.2$, the momentum distribution departs strongly from FL behavior and closely resembles the strongly violating $d_{xy}$-NFL phase of Fig.~\ref{fig:moment}(b).
In summary, the momentum distribution reveals two distinct mechanisms by which the LL theorem is violated: a \emph{weak} violation characterized by a continuous deformation of the Fermi surface and a gradual redistribution of spectral weight, and a \emph{strong} violation marked by abrupt shifts of the singular momenta and the disappearance of a well-defined Fermi surface. These two regimes correspond, respectively, to $d_{x^2-y^2}$- and $d_{xy}$-wave NFL phases, as we will show below by analyzing their pairing symmetries.

\begin{figure}[t]
    \centering
    \includegraphics[width=\columnwidth]{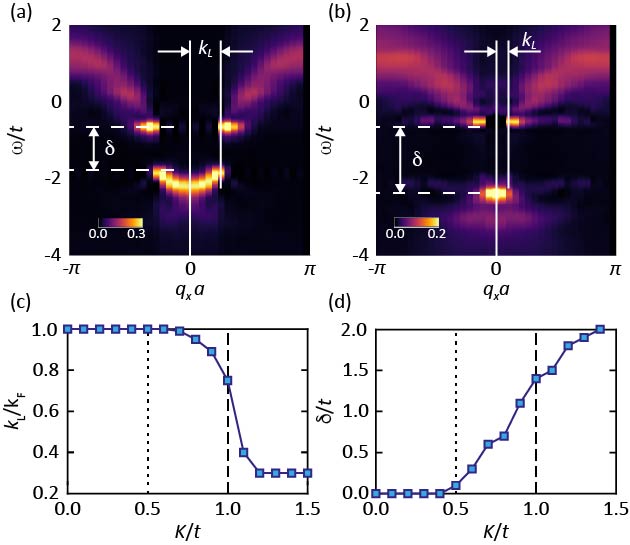}
    \caption{Spectral function analysis for $t_d=-0.25$, $q_y=0$, and $L_x=36$. Two representative values are shown in (a) $K=0.8$ and (b) $K=1.2$.
    (c) Gap $\delta$ and (d) corresponding momentum $k_L$, schematically shown in the (a)-(b), as a function of $K$. Vertical lines correspond to FL-NFL transition points.}
    \label{fig:SF}
\end{figure}

Transitions between different phases can also be identified using dynamical properties, providing us with not only additional characteristics of the NFL nature of these phases, but also data that are particularly important in light of recent advances in experimental methods such as INS, RIXS and ARPES~\cite{ARPES,RIXS}. These methods allow for the measurement of the response of quantum many-body systems at specific energies and momenta, and specifically of the spectral function (cf.~\cite{SM}), analyzed in Fig.~\ref{fig:SF}. Panels (a) and (b) show two examples of the spectral function in two distinct NFL regimes. Importantly, both regimes are characterized by a gap $\delta$ that opens at momentum $k_L<k_F$, approximately coinciding with the peak of the derivative $|\partial_{q_x} \la n_{\uparrow}(q_x,0) \ra|$ analyzed in Fig.~\ref{fig:moment}(c,f). 
The gap $\delta$ and the momentum $k_L$ as a function of $K$ are shown in Fig.~\ref{fig:SF} (c) and (d) for moderate diagonal hopping $t_d=-0.25$. We notice that the opening of a gap coincides with the onset of LL theorem violation, marked by the dotted horizontal lines. This behavior corresponds to the transition from the FL to NFL. We further notice a sharp drop in $k_L$ at $K\approx1$, marked by the vertical dashed line. This drop separates the $d_{x^2-y^2}$-wave NFL regime, in which the LL theorem is weakly violated, from the $d_{xy}$-wave NFL regime, characterized by strong NFL behavior. 
These differences are qualitatively observed in panels (a) and (b) of Fig.~\ref{fig:SF}: While in Fig.~\ref{fig:SF}(a) we observe a gapped phase with well-defined dispersion, corresponding to the (weak) $d_{x^2-y^2}$-wave NFL phase, Fig.~\ref{fig:SF}(b) shows a more complex structure, characteristic for the (strong) $d_{xy}$-wave NFL phase. 

\begin{figure}
    \centering
    \includegraphics[width=\columnwidth]{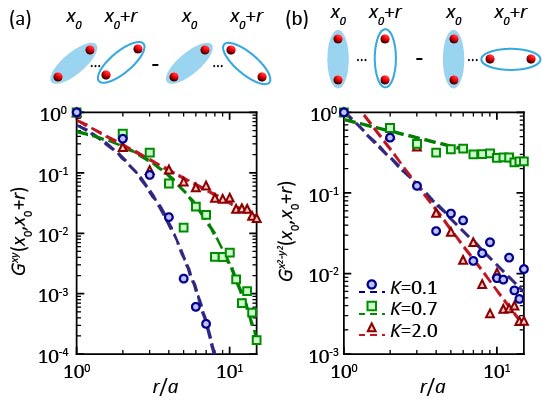}
    \caption{Pair correlations for the two $d$-wave channels for $t_d=-0.25$, $L_x=30$, and three distinct values of $K$, representative of the three phases. (a) Diagonal (alias $d_{xy}$) and (b) axial (alias $d_{x^2-y^2}$) $d$-wave pair correlations, as schematically illustrated in the top panels. Dashed curves represent the best numerical fit between power-law and exponential. Correlations are renormalized to the first point to better illustrate the decay laws.}
    \label{fig:dwave}
\end{figure}

\textit{Pairing correlations.}
With the complementary observables introduced so far, we have differentiated between three phases: FL, weak NFL, and strong NFL. As we show in the following, the distinction between these phases, and in particular between the different NFL regions, becomes clearest if we consider pairing correlations.
In the absence of the band flattening term, the NFL phase has been identified as $d$-wave because of the $d$-wave symmetry exhibited by its Cooper-pair correlations~\cite{jiang2013non}. As we show below, different pairing channels become relevant in the model with diagonal hopping. To identify the type of pairing, we analyze the singlet-pairing density matrix. 
In general, this quantity reads
\begin{equation}\label{eq:density_matrix}
    \rho_S(\bm{r}_i,\bm{r}_j|\bm{r}_k,\bm{r}_l)=\langle \hat{\Delta}^\dagger_{ij} \hat{\Delta}_{kl}\rangle.
\end{equation}
To exclude local contributions from density and spin correlations, we set the density matrix $\rho_S(\bm{r}_i|\bm{r}_j)$ to zero whenever there is overlap of Cooper pairs that share a site~\cite{Wietek2022}. Therefore, $\rho_S$ is no longer positive definite, so some of its eigenvalues may be negative.
For the $d$-wave pairing, two channels are taken into account in our geometry:
\begin{subequations}
\begin{align}
    G_{ij}^{\:xy}=&\:\rho_S(\bm{r}_i,\bm{r}_i+\hat{\bm x}+\hat{\bm y}|\bm{r}_j,\bm{r}_j+\hat{\bm x}+\hat{\bm y})\nonumber \label{eq:Gxy}\\ 
    - &\:\rho_S(\bm{r}_i,\bm{r}_i+\hat{\bm x}+\hat{\bm y}|\bm{r}_j+\hat{\bm y},\bm{r}_j+\hat{\bm x}),
    \\
    G_{ij}^{\:x^2-y2}=&\:\rho_S(\bm{r}_i,\bm{r}_i+\hat{\bm y}|\bm{r}_j,\bm{r}_j+\hat{\bm y})\nonumber \label{eq:Gx2y2}\\ 
    - &\:\rho_S(\bm{r}_i,\bm{r}_i+\hat{\bm y}|\bm{r}_j,\bm{r}_j+\hat{\bm x}),
\end{align}
\end{subequations}
with the first representing \textit{diagonal} $d$-wave ($xy$) pairs while the second depicts \textit{axial} $d$-wave ($x^2$--$y^2$) pairs. We show a visual representation in Fig.~\ref{fig:dwave} and plot the correlations as a function  of distance $r=|\bm{r}_i - \bm{r}_j|$ within the bulk. In this figure, $t_d=-0.25$ is fixed, and we choose three values of $K$ as representatives of the three regimes identified earlier (Fermi-liquid and two NFL phases), allowing a direct visual comparison between the character of pair correlations across phases. Fig.~\ref{fig:dwave}(a) shows the diagonal $d_{xy}$-wave correlations, which change qualitatively from an exponential decay at small and moderate $K$ to a clear algebraic tail at strong ring exchange interaction, signaling the emergence of quasi-long-range order in the strong-NFL region. 
In contrast, Fig.~\ref{fig:dwave}(b) displays the axial $d_{x^2-y^2}$ channel, where the correlations show a slow, almost flat, decay with distance $r$ for moderate $K$, while a faster than polynomial decay is observed in both the FL and strong-NFL regimes.
The dashed lines represent the best numerical fits distinguishing the exponential and power-law decay laws. Although blurred by finite-size effects, a clear distinction is observed between short- and long-range correlations.

A clear theoretical signature of fragmented Cooper-pair condensation is provided by the generalized Penrose--Onsager criterion \cite{Penrose-Onsager,leggett2006quantum}. The emergence of Cooper-pair condensation in a given pairing channel is signaled by a dominant eigenvalue $\varepsilon_n$ of the corresponding pairing density matrix,
$\rho_S(\bm{r}_i|\bm{r}_j)=\sum_n \varepsilon_n \chi_n^*(\bm{r}_i)\chi_n(\bm{r}_j)$,
which directly encodes the behavior of $G_{ij}^\xi$. The eigenvalue gap $\varepsilon_0-\varepsilon_1$ therefore serves as a sensitive diagnostic for identifying pairing symmetry, as illustrated in Fig.~\ref{fig:phase_diagram}. We find that these gaps remain robust upon increasing system size (see Supplemental Material~\cite{SM}), where we also present the full eigenvalue spectra across the different phases.

\begin{figure}[t]
    \centering
    \includegraphics[width=0.95\linewidth]{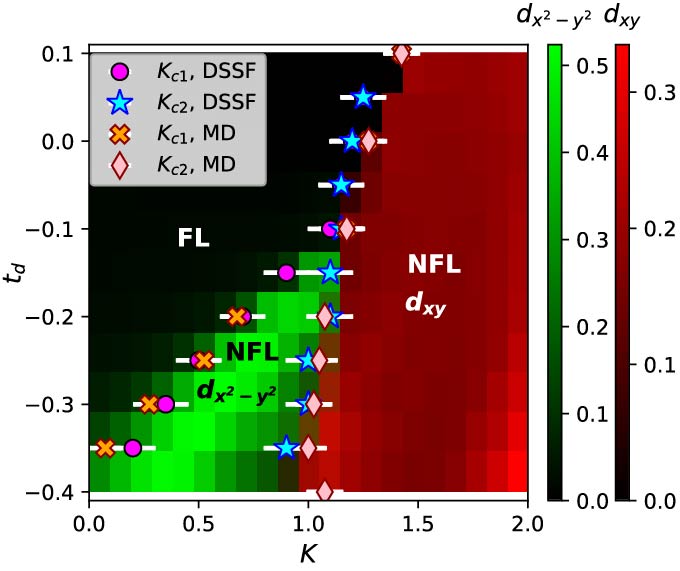}
    \caption{Finite-size phase diagram depicting the three main phases observed: a Fermi-liquid phase (FL) and two  non-Fermi-liquid (NFL) phases with distinct $d$-wave pairing symmetries. The color background shows the eigenvalue gap in the dominant pairing channel, i.e. $d_{x^2-y^2}$ (green) or $d_{xy}$ (red). As indicated by
    the symbols, the phase boundaries can also be obtained from inspection of the momentum distribution (MD) and the dynamical spin structure factor (DSSF), where critical values $K_{c1}$ and $K_{c2}$ mark the transitions between the different phases. The grid of $K$ values in the calculation yields horizontal error bars.}
    \label{fig:phase_diagram}
\end{figure}

\textit{Implementation. }
In physical systems, ring-exchange interactions can arise from nonlocal interaction potentials. In neutral-atom quantum simulation experiments, such processes typically emerge only at higher order, as demonstrated, for example, in Ref.~\cite{dai_four-body_2017}. Identifying schemes that enhance the energy scale of ring-exchange interactions has long been a central challenge in quantum simulation, particularly in the context of lattice gauge theories~\cite{feldmeier_quantum_2024}. In contrast, sizeable negative diagonal hopping amplitudes appear to be directly accessible with state-of-the-art Floquet engineering techniques or by exploiting synthetic dimensions (cf. Refs.~\onlinecite{Goldman2014,celi_synthetic_2014,Grass2015,Eckhardt2017}). In particular, diagonal tunneling couplings with tunable complex amplitudes have been demonstrated in simulations of Creutz ladders~\cite{kang_creutz_2020} and in Raman-coupled lattices~\cite{song_observation_2018}. We note, however, that engineering the required interaction terms is more challenging in approaches based on synthetic dimensions. A well-known limitation of Floquet systems is heating, especially in the presence of interactions, which has so far constrained progress toward realizing strongly correlated topological phases of matter. Recent experiments, however, have reported the realization of large Floquet many-body systems of strongly interacting bosonic atoms subject to artificial gauge fields at temperatures on the order of the tunneling energy~\cite{Impertro2025}, underscoring the experimental feasibility of our proposed scheme.  To assess on the thermal stability of the $d_{x^2-y^2}$ pairing phase at these temperatures, we calculated the pairing gap, see SM \cite{SM}, and show that a gap larger than $0.5t$ can be achieved. Importantly, one of these ingredients -- either diagonal hopping or ring exchange -- is sufficient to drive a transition from a FL into a NFL $d$-wave pairing phase, hence our manuscript outlines two different avenues to achieve the outstanding goal of realizing and controlling pairing physics in Fermi-Hubbard systems and directly connects to recent experimental efforts in engineering low-entropy phases in the Fermi-Hubbard model with next-nearest-neighbor tunneling~\cite{xu_neutral-atom_2025}. 

\textit{Summary.}
We find that ring-exchange interactions combined with negative diagonal hopping in a strongly repulsive Fermi–Hubbard ladder stabilize distinct non-Fermi-liquid phases, characterized by the breakdown of Luttinger liquid theory, quasi-long-range $d$-wave–like correlations, and fragmented Cooper-pair condensation. Our minimal model elucidates how band flattening generates competition among pairing symmetries. Identifying diagonal tunneling as an efficient route to stabilize $d$-wave pairing, our results open new perspectives for engineered synthetic matter systems.

\textit{Acknowledgments.}
We thank Ravindra Chhajlany and Tymoteusz Salamon for the fruitful discussions in the early stages of the project.
J.P.M. and K.J. were supported by the Polish National Agency for Academic Exchange (NAWA) via the Polish Returns 2019 program and gratefully acknowledge Polish high-performance computing infrastructure PLGrid (HPC Center: ACK Cyfronet AGH) for providing computer facilities and support within computational grant no. PLG/2024/017478. Part of the numerical calculations have also been carried out using the resources provided by the Wrocław Centre for Networking and Supercomputing.
The "Quantum Optical Technologies" (FENG.02.01-IP.05-0017/23) project is carried out within the Measure 2.1 International Research Agendas programme of the Foundation for Polish Science, co-financed by the European Union under the European Funds for Smart Economy 2021-2027 (FENG).
M.M.M. acknowledges support from the National Science Centre (Poland) under Grant No. 2024/53/B/ST3/02756. M.D. acknowledges support from the National Science Centre (Poland) under Grant No. 2022/04/Y/ST3/00061.
M.A. acknowledges support from the Deutsche Forschungsgemeinschaft (DFG, German Research Foundation) under Germany’s Excellence Strategy -- EXC-2111 -- 390814868 and the Horizon Europe programme HORIZON-CL4-2022-QUANTUM-02-SGA via the project 101113690 (PASQuanS2.1).
T.G. acknowledges funding by the Department of Education of the Basque Government through the IKUR Strategy, through BasQ (project EMISGALA), and through PIBA\_2023\_1\_0021 (TENINT), and by the Agencia Estatal de Investigación (AEI) through Proyectos de Generación de Conocimiento PID2022-142308NA-I00 (EXQUSMI). This work has been produced with the support of a 2023 Leonardo Grant for Researchers in Physics, BBVA Foundation. The BBVA Foundation is not responsible for the opinions, comments, and contents included in the project and/or the results derived therefrom, which are the total and absolute responsibility of the authors.

\end{document}


\title{Supplemental Material to \\ ``Controlled pairing symmetries in a Fermi-Hubbard ladder with band flattening''}

\author{Jo\~{a}o P. Mendon\c{c}a} 
\email{jpedromend@gmail.com}
\affiliation{\affCent}
\affiliation{\affFUW}
\author{S. Biswas}
\affiliation{\affDIPC}
\author{M. Dziurawiec}
\affiliation{\affWroclaw}
\author{U. Bhattacharya}
\affiliation{Institute for Theoretical Physics, ETH Z\"urich, Z\"urich 8093, Switzerland}
\author{K.~Jachymski}
\affiliation{\affFUW}
\author{M. Aidelsburger}
\affiliation{Max-Planck-Institut f\"ur Quantenoptik, 85748 Garching, Germany}
\affiliation{Munich Center for Quantum Science and Technology (MCQST), 80799 M\"unchen, Germany}
\affiliation{Fakult\"at f\"ur Physik, Ludwig-Maximilians-Universit\"at, 80799 M\"unchen, Germany}
\author{M. Lewenstein}
\affiliation{\affICFO}
\author{M. M. Maśka}
\affiliation{\affWroclaw}
\author{T. Grass}
\email{tobias.grass@dipc.org}
\affiliation{\affDIPC}
\affiliation{\affIker}

\date{\today}
\maketitle

We provide additional information related to dynamical properties (spectral function, dynamical spin structure factor), to the generalized Penrose-Onsager criterion, obtained from the eigenvalue spectrum of the non-local $d$-wave density matrices, and on the finite-size scaling of the eigenvalue gaps. 

\section{Spectral function}
The single-particle spectral function $A({\bm q},\omega)$ can be obtained by computing the nonequal time two-point correlators via a real-time evolution. To this end, we time evolve the ground state through a local quench, from which the dynamical correlation functions can be extracted at each time step. Thus,
\begin{align}
    A_\sigma({\bm q},\omega) &= \sum_{j}e^{-i{\bm q}\cdot {\bm r}_j}\nonumber \\
    &\times \dfrac{1}{2\pi}\left[
    \int_{-\infty}^\infty dt\: e^{i(\omega-E_0)t} \langle c_{j\sigma}(t)c^\dagger_{0\sigma}(0) \rangle\right.\nonumber \\
 & + \left.\int_{-\infty}^\infty dt\: e^{-i(\omega+E_0)t} \langle  c^\dagger_{j\sigma}(t)c_{0\sigma}(0) \rangle\right],
\end{align}
where the expectations are taken in the ground state whose energy is $E_0$. Here, we performed the time evolution using the time-dependent variational principle method \cite{Dirac_1930,PhysRevLett.107.070601}.

\section{Dynamical spin structure factor.}
\begin{figure}[t]
    \centering
    \includegraphics[height=0.53\linewidth]{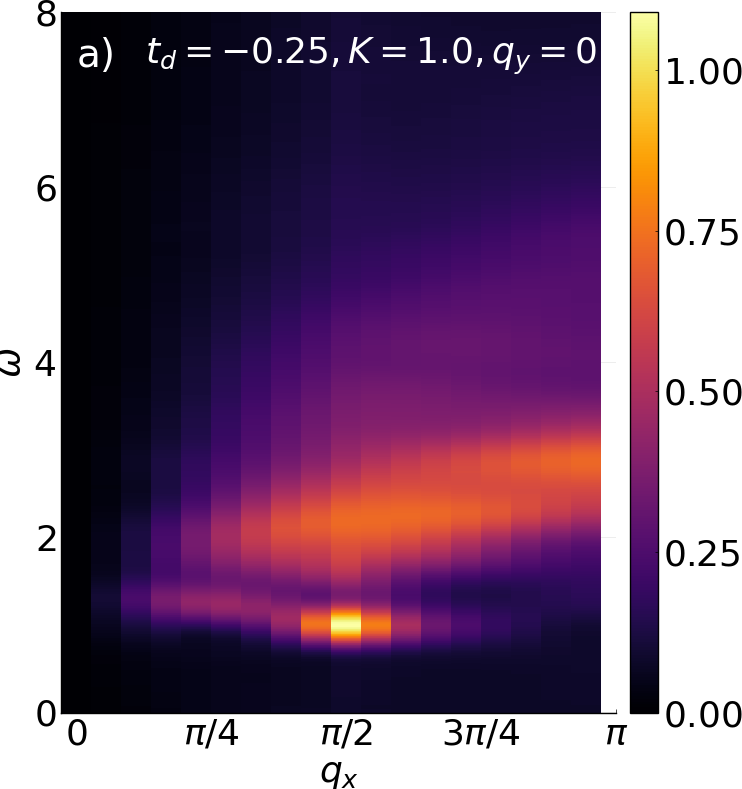}\hspace*{1mm}\includegraphics[height=0.53\linewidth]{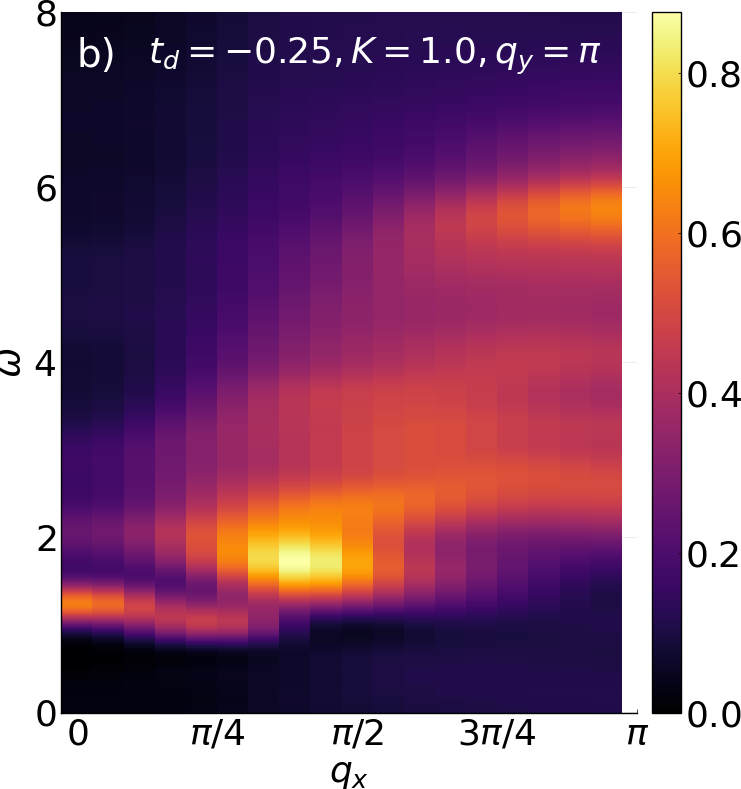}\\[1mm]
    \includegraphics[height=0.53\linewidth]{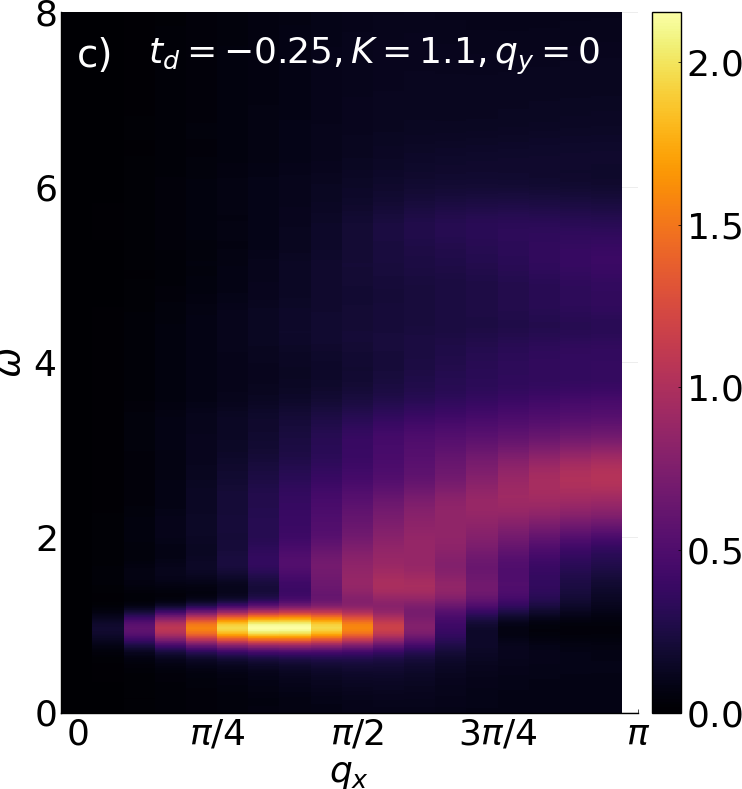}\hspace*{1mm}\includegraphics[height=0.53\linewidth]{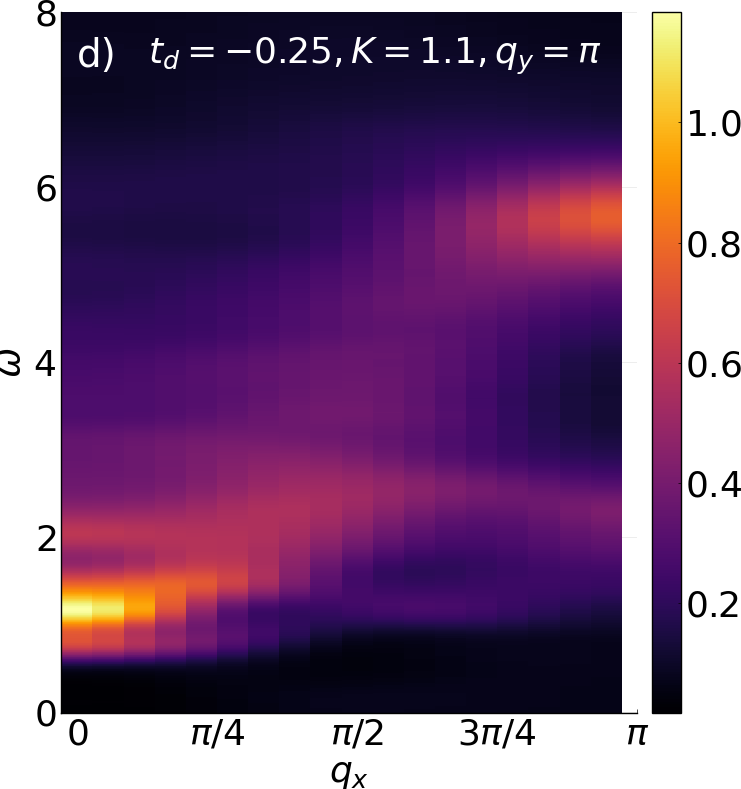}
    \caption{Dynamical spin structure factor $S^{zz}({\bm q},\omega)$ for $t_d=-0.25$, $K=1.0$ [(a) and (b)] and $K=1.1$ [(c) and (d)]. Results are presented for $q_y=0$ [(a) and (c)] and $q_y=\pi$ [(b) and (d)].}
    \label{fig:DSSF}
\end{figure}
The dynamical spin structure factor $S^{zz}({\bm q},\omega)$ is written as
\begin{equation}
    S^{zz}({\bm q},\omega) = \sum_{j}e^{-i{\bm q}\cdot {\bm r}_j} \int_{-\infty}^\infty dt\: e^{i\omega t}
    \langle S^z_j(t)S^z_0(0) \rangle.
\end{equation}
Figure \ref{fig:DSSF} shows an example of how $S^{zz}({\bm q},\omega)$ changes when we go from $K=1.0$ to $K=1.1$, that is, when the system goes from the $d_{x^2-y^2}$-wave phase to the $d_{xy}$-wave phase. Clearly, for $q_y = \pi$, the maximum weight of $S^{zz}({\bm q}, \omega)$ abruptly shifts from a finite $q_x$ to $q_x = 0$. Similar changes can also be observed for other transitions, which were used as additional indicators of phase transitions.

\section{Generalized Penrose-Onsager criterion.}
\begin{figure}[t]
    \centering
    \includegraphics[width=0.99\columnwidth]{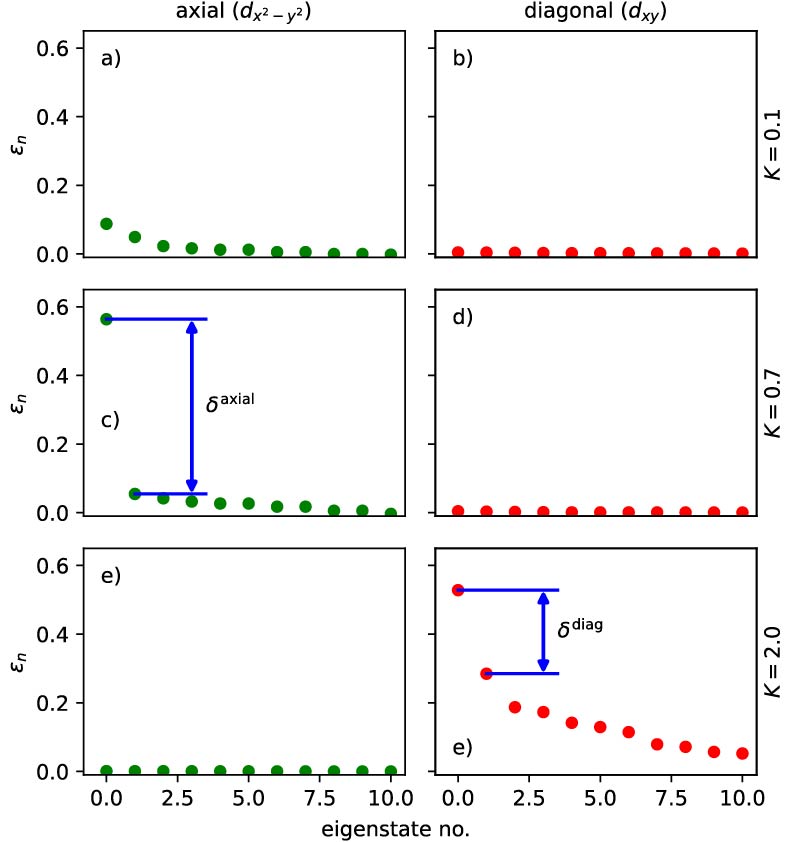}
    \caption{The lowest 10 eigenvalues of the density matrices $G^{x^2-y^2}$ [(a), (c), (e)] and $G^{xy}$ [(b), (d), (f)] [cf. Eqs.~(3a) and (3b) for $t_d=-0.25$ and $K=0.1$ [(a), (b)], $K=0.7$ [(c), (d)], and $K=2.0$ [(e), (f)]. The system size is $30\times 2$.}
    \label{fig:eivals}
\end{figure}
\vfill\null
We show in Fig.~\ref{fig:eivals} the highest eigenvalues of the $d$-wave correlations, given by Eqs.~(3a) and (3b). In all panels, we set $t_d=-0.25$ and choose
$K=0.1$ in panel (a,b), $0.7$ in panel (c,d), and $2.0$ in panel (e,f). These parameters correspond to the three different phases observed in our model. For small $K$, panels (a) and (b), no eigenvalue dominates the spectrum, and hence, neither diagonal nor axial pairs are condensed. In contrast, at an intermediate $K$, panels (c) and (d), a pronounced eigenvalue emerges in the eigenvalue spectrum of the axial $d_{x^2-y^2}$-wave pairing. Finally, for a strong $K$, panels(e) and (f), a dominant eigenvalue appears in the diagonal $d_{xy}$-wave pairing correlations. Consequently, the eigenvalue gap between the largest and second-largest, denoted $\delta^{\rm axial}$ and $\delta^{\rm diag}$, respectively, can be used as order parameters of the different pairing phases, as also done in Fig.~5 of the main text.

\section{Effect of finite system size.} 
\begin{figure}[!h]
    \centering
    \includegraphics[width=0.99\columnwidth]{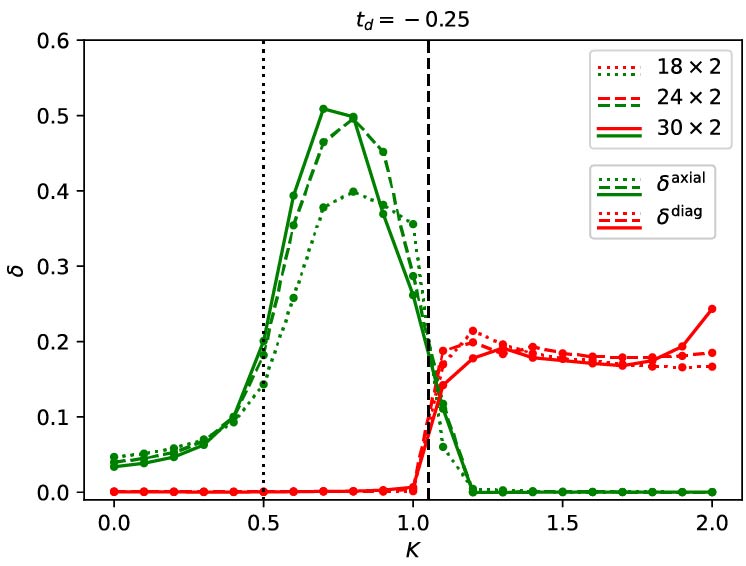}
    \caption{The magnitude of the eigenvalue gap for $18\times 2$ (dotted lines), $24\times 2$ (dashed lines), and $30\times 2$ (solid lines) systems for $t_d=-0.25$. The green lines show $\delta^\mathrm{axial}$ and the red lines show $\delta^\mathrm{diag}$.}
    \label{fig:fss}
\end{figure}

The finite-size analysis of the gaps shown in Fig.~\ref{fig:fss}. A comparison of the magnitude of the eigenvalue gap for systems of different sizes indicates that the systems used in the calculations presented in this work ($36\times 2$ for the spectral functions and dynamical spin structure factors and $30\times 2$ for the other quantities) are large enough to determine the phase boundaries within the error bars marked in Fig.~5.

\section{Pairing gap}
\begin{figure}[t]
    \centering
    \includegraphics[width=0.99\columnwidth]{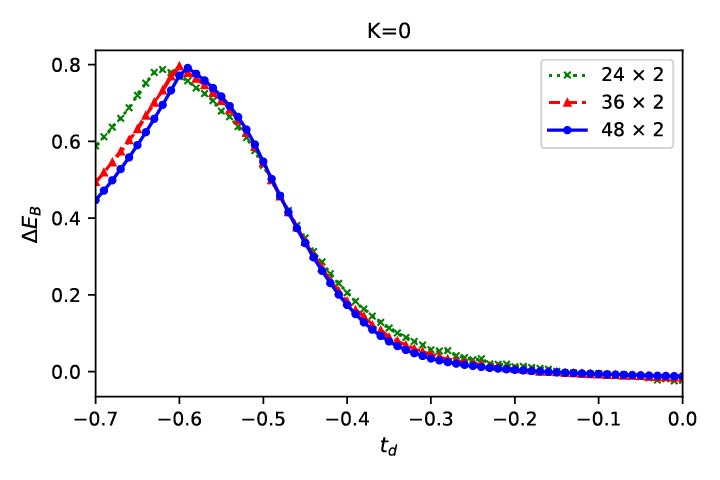}
    \caption{The magnitude of the binding energy \eqref{EB}  for three system sizes $L_x = 24$ (green), $L_x=36$ (red), $L_x=48$ (blue) at $K=0$.}
    \label{fig:pairing_gap}
\end{figure}
In order to estimate the magnitude of the pairing gap, we calculate the binding energy of a pair of electrons,
\begin{equation}
\label{EB}
    \Delta E_B = 2E_0(N_e+1) - E_0(N_e+2) - E_0(N_e),
\end{equation}
where $E_0(N)$ is the energy of the groundstate in a system with $N$ particles and $N_e$ is the number of particles corresponding to the filling $\rho =1/3$ considered in this work. It can be intuitively understood as the energy cost difference of adding two independent electrons instead of a pair. In Fig.~\ref{fig:pairing_gap} we show that $\Delta E_B$ indicates gap opening in the axial $d_{x^2-y^2}$-wave pairing phase.

\bibliography{refs}
\vspace{26mm}